\documentclass[11pt]{article}
\usepackage{amsxtra,amssymb,amsfonts,graphics}
\begin{document}
\title{ACTIVE MASS UNDER PRESSURE}
\author{J. Ehlers, I. Ozsv\'{a}th and E.L. Schucking}
\date{May 2, 2005}
\maketitle
\begin{abstract}
After a historical introduction to Poisson's equation for Newtonian gravity,
its analog for static gravitational fields in Einstein's theory is reviewed. It
appears that the pressure contribution to the active mass density in Einstein's
theory might also be noticeable at the Newtonian level. A form of its 
surprising appearance, first noticed by Richard Chase Tolman, was discussed 
half a century ago in the Hamburg Relativity Seminar and is resolved here.
\end{abstract}

\section{Historical Introduction}\label{S:i}
\setcounter{equation}0

By ``active'' mass one means the mass considered as source of a gravitational
field. This concept can be distinguished from that of the ``passive'' mass 
measuring the response to a gravitational field and of an ``inertial'' mass 
describing resistance against acceleration by any force, gravitational or 
other. These distinctions are useful for considering non-Newtonian or 
non-Einsteinian theories of gravity, or to assess which aspects of mass are 
tested in experiments. In Newton's and Einstein's theories these distinctions 
are not necessary, in agreement with available physical and astronomical 
evidence and post-Newtonian approximations.

Joseph Louis Lagrange in his 1773 ``Memoir on the secular Equation of the 
Moon'' \cite{JLL} introduced the function $V$ through the equation
\begin{equation}\label{i.1}
V(\vec{x}) = \int \frac{\rho(\vec{x'})}{|\vec{x}-\vec{x'}|}d^3\vec{x'}.
\end{equation}

He recognized that it is easier to first compute $V$ and then to get the 
accelerations by differentiation than to calculate the accelerations directly.

The function $\rho(\vec{x'})$ denotes here the density of the active mass. The
active mass is then given by
\begin{equation}\label{i.2}
M = \int\rho(\vec{x'})d^3\vec{x'}.
\end{equation}

Lagrange did not use vector notation and did not exhibit Newton's
gravitational constant $G$ because astronomers set it equal to one as we
do in this paper. We recognize now Lagrange's function $V$ as the negative of 
the gravitational potential $\phi$. The use of the negative of the potential 
was customary before the principle of the conservation of energy began to 
dominate physics and astronomy.

Lagrange did not supply a name for his function $V$. Carl Friedrich Gauss 
called it the potential in his 1839 paper on ``General Theorems about Forces 
of Attraction and Repulsion depending on the inverse Square of the Distance''
\cite{CFG}. Gauss had not been aware of George Green's 1828 article ``An
Essay on the Application of Mathematical Analysis on the Theory of Electricity
 and Magnetism'' \cite{GG} that had named $V$ the ``potential function''. 
Green had published his paper privately and had rarely referred to it in his 
later papers. It was only several years after Green's death in 1841 that 
William Thomson (the later Lord Kelvin) discovered Green's paper and arranged
for the publication of its results.

The acceleration vector $\ddot{\vec{x}}$ for a massive particle was given by 
Lagrange in an inertial system by
\begin{equation}\label{i.3}
\ddot{\vec{x}} = \bigtriangledown V.
\end{equation}

This law may be read as saying that the inertial mass may be identified with 
the passive gravitational mass.
 
In his 1782 ``Theory of the attraction of spheroids and the figure of the 
earth'' \cite{PL1} Pierre Simon Laplace introduced for the function $V$ the
equation
\begin{equation}\label{i.4}
\bigtriangledown^2 V \equiv \frac{\partial^2 V}{\partial x^2} + \frac{\partial^2 V}{\partial y^2} + \frac{\partial^2 V}{\partial z^2} = 0
\end{equation}

that is now named the Laplace equation. Laplace did not write it, as we just 
did, in Cartesian coordinates $x, y, z$ but used spherical polar coordinates.
He published the form in  Cartesian coordinates in his ``Memoir about the 
theory  of Saturn's ring''  \cite{PL2} in 1787.

It was only 26 years later that Sim\'{e}on-Denis Poisson pointed out in his
``Remarks about an equation that occurs in the theory of attraction of  
spheroids'' that the Laplace equation does not hold within the substance of 
the attracting body \cite{SDP} and has there to be replaced by
\begin{equation}\label{i.5}
\bigtriangledown^2 V = -4\pi \rho.
\end{equation}

This equation is now known as the Poisson equation. A proof 
 was first given by Gauss in 1839 \cite{CFG}. If we write the 
equations in terms of the potential $\phi$ we have
\begin{equation}\label{i.6}
\bigtriangledown^2 \phi = 4\pi \rho,
\end{equation}

the relation between the gravitational potential and the density of the active 
mass in an inertial system of Newton's theory. If $\phi$ is required to vanish
at infinity, (\ref{i.6}) implies (\ref{i.1}). 

\section{Static Fields in Einstein's Theory of Gravitation}\label{S:gr}
\setcounter{equation}0

In General Relativity \cite{GR} the ten components $g_{\mu\nu}$ of the metric tensor
\begin{equation}\label{gr.1}
ds^2 = g_{\mu\nu}(x^\lambda)dx^\mu dx^\nu
\end{equation}

replace the potential of Lagrange. Moreover, not just the mass-density, but all
ten components of the energy-momentum-stress tensor $T_{\mu\nu}$ become 
contributing sources to the gravitational field. For neutral matter 
$T_{\mu\nu}$ is given by
\begin{equation}\label{gr.2}
T_{\mu\nu} = \rho u_\mu u_\nu + p_{\mu\nu}
\end{equation}

with energy density $\rho$, pressure tensor $p_{\mu\nu}$ and four-velocity 
$u^\mu$ subject to the normalisation 
\begin{equation}\label{gr.3}
u^\mu u_\mu = 1,
\end{equation}

and
\begin{equation}\label{gr.4}
p_{\mu\nu} u^\nu = 0.
\end{equation}

For a perfect fluid
\begin{equation}\label{gr.5}
p_{\mu\nu} = p(u_\mu\,u_\nu - g_{\mu\nu}).
\end{equation}

Instead of Poisson's equation relating the gravitational potential to the 
active mass density $\rho$ we have now the Einstein field equations
\begin{equation}\label{gr.6}
R_{\mu\nu} - \frac{1}{2}g_{\mu\nu}R + \lambda g_{\mu\nu}  = - 8\pi T_{\mu\nu}
\end{equation}

connecting the $g_{\mu\nu}$ and their derivatives up to the second order to
the components of $T_{\mu\nu}$. The Riemann tensor is defined through the 
interchange of order in the second covariant derivatives of a covariant vector
field $\xi_\mu$
\begin{equation}\label{gr.7}
\xi_{\mu;\alpha;\nu} - \xi_{\mu;\nu;\alpha} = - \xi_\gamma R^\gamma{}_{\mu\alpha\nu}
\end{equation}

and the Ricci tensor by
\begin{equation}\label{gr.8}
R_{\mu\nu} = R^\gamma{}_{\mu\gamma\nu}.
\end{equation}

The Ricci scalar $R$ is obtained through contraction from the Ricci tensor 
$R^\mu{}_\nu$.

Here we use units where the speed of light ``c'' is put equal to 1.

For a static gravitational field in which the matter is at rest we are able to
find an analog of Lagrange's potential and its Poisson equation. In 
section~\ref{S:a1} 
we give an invariant derivation of the necessary developments. Here we give 
just the results. A static gravitational field can be described by the metric
\begin{equation}\label{gr.9}
ds^2 = g_{00}(x^l)(dx^0)^2 + g_{jk}(x^l)dx^j dx^k
\end{equation}

with no $g_{0i}$ terms. Latin indices run here from 1 to 3. The components of
the metric tensor do not depend on the time coordinate $t = x^0$ and the four- 
velocity $u^\mu$ of the matter is given by
\begin{equation}\label{gr.10}
u^\mu = \frac{1}{\sqrt{g_{00}}}\,\delta^\mu_0.
\end{equation}

We have then for the static potential $\sqrt{g_{00}}$ with 
$\hat{g} = \det\parallel g_{jk} \parallel$
\begin{equation}\label{gr.11}
-\frac{1}{\sqrt{-\hat{g}}}\Big(\sqrt{-\hat{g}}\,g^{jk}\,(\sqrt{g_{00}})_{,k}\Big)_{,j} \equiv \bigtriangledown^2\sqrt{g_{00}} = 4\pi\sqrt{g_{00}}\Big(\rho + 3p - \frac{\lambda}{4\pi}\Big).
\end{equation}

The spatial metric differs from the flat metric only by terms of order $\phi$,
On the left-hand side of the relativistic Poisson equation we have the Laplace
operator on the spaces t = const. for the function $\sqrt{g_{00}}$ that we now 
identify as the static relativistic potential. This expression in Riemannian
geometry was derived by Eugenio Beltrami in his 1868 paper ``On the general 
theory
of differential parameters'' and became known as the second Beltrami parameter
 \cite{EB}.

The cosmological term in (\ref{gr.6}) can be put on the right-hand side of the equation as
\begin{equation}\label{gr 12}
- \lambda g_{\mu\nu} \equiv - 8\pi T_{\mu\nu}(\lambda),
\end{equation} 

Interpreted as an energy-momentum tensor of a perfect fluid 

\begin{equation}\label{gr 13}
T_{\mu\nu}(\lambda) = [\rho(\lambda) + p(\lambda)]u_\mu u_\nu -p(\lambda)g_{\mu\nu}
\end{equation}
we have
\begin{equation}\label{gr 14}
\rho(\lambda) = - p(\lambda), \quad p(\lambda) = - \frac{\lambda}{8\pi}.
\end{equation}

The contribution to the active mass density (``dark energy'') becomes

\begin{equation}\label{gr 15}
\rho(\lambda) + 3 p(\lambda) = - \frac{\lambda}{4\pi}.
\end{equation}

On the right-hand side we have Poisson's $4\pi \rho$
term. Apart from Einstein's lambda term that acts now--if positive--as a 
negative mass density contributing to the active mass,  we have two changes
in comparison with Poisson's equation. One is the factor $\sqrt{g_{00}}$ on
the right hand side of the equation. Since the discovery of energy 
conservation the potential had
been defined as the work to displace a unit mass (charge) from infinity and it
was usually normalized to vanish at spatial infinity. In Einstein's theory of
gravitation the component $g_{00}$ for a finite mass distribution in a static 
gravitational field is usually taken as $c^2$ or 1 at infinity. The 
reason is that now one includes the rest energy of a particle as part of the 
potential energy and assumes that the metric of an isolated system becomes 
Euclidean at large distances. We have then for $\sqrt{g_{00}}$ at large 
distances $r$ from the center of mass
\begin{equation}\label{gr.16}
\sqrt{g_{00}}(r) \approx 1 - \frac{M}{r}
\end{equation}

where $M$ is the total active mass. Thus, for weak fields the factor 
$\sqrt{g_{00}}$ in the relativistic Poisson equation differs from 1 only
slightly. 

The most surprising correction to the relativistic Poisson equation is the term
$3p$ that is actually $3p/c^2$. This term was first noted explicitly as a
consequence of Einstein's field equations by Tullio Levi-Civit\'{a} in 1917 in
his series of papers on ``Einstein's Static'' \cite{TLC}. It is this term that
we shall discuss in other sections of this paper. Finally, 
$\sqrt{\hat{g}}\,d^3\vec{x'}$ in the formula for the active mass is the 
proper volume element.

The justification for identifying $\sqrt{g_{00}}$ with the relativistic 
global potential rests on its definition as the specific potential energy. We
shall show in section~\ref{S:eps} that the test particle of unit mass resting 
at $\vec{x}$ has potential energy (apart from its rest-energy)
\begin{equation}\label{gr.17}
\phi = \sqrt{g_{00}} - 1.
\end{equation}

The Poisson equation in Newtonian gravity is written down in a global inertial
system. The closest analog to such a system for Einstein's static gravitational
fields (without cosmological constant) is based on the coordinates adapted to
the time-like hypersurface-orthogonal Killing vector. It also defines a rest 
system through the divergence-free time-like unit vectors orthogonal to the 
hypersurfaces t = const. reaching out to spatial infinity where the space-time
metric is assumed to become Minkowskian. These coordinates, it has to be
stressed, are \emph{not} local inertial systems. The acceleration of a 
particle at rest is given by
\begin{equation}\label{gr.18}
\dot{u}_k = -\,\frac{1}{\sqrt{g_{00}}}\frac{\partial}{\partial x^k}\sqrt{g_{00}} = - \frac{\phi_{,k}}{(1+\phi)}.   
\end{equation}

\section{Tolman's Paradox}\label{S:tp}
\setcounter{equation}0

Richard Chase Tolman in his 1930 paper ``On the use of the energy-momentum
principle in general relativity'' \cite{RCT} derived a formula for the total 
energy of a fluid sphere in a quasi-static state for Einstein's theory of 
gravitation. He wrote with a proper volume element $dV_0$
\begin{equation}\label{tp.1}
M = \int(\rho + 3p)\sqrt{g_{00}}\,dV_0 = \int\partial_a\phi\,d\vec{F}^a  
\end{equation}

suppressing the cosmological term, where $\rho$ is the relativistic energy 
density.

In statistical mechanics of an ideal gas consisting of particles with mass $m$,
and momentum $\vec{p}$, velocity $\vec{v}$, and number density $n$, the pressure $p$ 
is given by Daniel Bernoulli's formula \cite{DB} from his 1738 ``Hydrodynamics, commentaries about forces and motions in fluids''
\begin{equation}\label{tp.2}
p = \frac{1}{3}\overline{n\,\vec{p} \cdot \vec{v}}
\end{equation}

where the bar indicates averaging. In (\ref{tp.2}) $p$ is the kinetic 
contribution to the pressure. In general $p$, or $p_{jk}$ contains 
contributions from short-range interactions, which have to be added to
(\ref{tp.2}). Attractive interactions contribute negative $p$. This formula, 
as written above, holds also for a relativistic ideal gas as discussed by
 Franz Juettner in his paper on 
``Maxwell's law of velocity distribution in the relative theory''\cite{FJ}. 
In the high energy limit of photons with energy $\epsilon$ this gives
\begin{equation}\label{tp.3}
\vec{p} \cdot \vec{v} = \epsilon, \quad p = \frac{1}{3}\overline{n\,\epsilon} =
\frac{1}{3}\rho.
\end{equation}

Since Tolman was particularly interested in the ``gravitational mass of 
disordered radiation,'' the $3p$-term in the active mass density caught his
attention.
This was no longer one of the tiny effects of general relativity. Tolman's 
observation led to the following paradox: while matter at rest in a spherical 
container of negligible matter might exhibit a total mass $M$ in the far field
of the container, transformation of the mass inside the container into 
disordered radiation would then double the total mass violating the 
conservation of
$M$. After the discovery of the positron in 1932 and its annihilation with the
electron such complete transformations of matter into radiation were no longer
impossible.

\section{The Hamburg paradox}\label{S:hp}
\setcounter{equation}0

The authors of this paper were members of a seminar on relativity that 
regularly met at Hamburg University in the 1950's. When we learned about the
$3p$-term in the Poisson equation the following test was suggested:

Since nucleons move in atomic nuclei with about two tenths of the speed of 
light, the $3p$-term might significantly contribute to the active mass density
 of  all
nuclei except hydrogen and the neutron. A simple calculation for the pressure 
in an ideal Fermi gas of nucleons at zero temperature (see section~\ref{S:b1})
gave a
pressure contribution to the active mass density of 4.3\,\% for nuclear matter.
A ball of hydrogen should, therefore, have an active mass about 4\,\% smaller 
then a ball of lead of the same inertial and passive gravitational mass that
could be checked by weighing them on scales.

However, the only way such an effect might be seen in the laboratory was by the 
Cavendish experiment for the determination of the gravitational constant where
 the active mass came into play. While it would be forbidding to work with a
ball of ultra-cold solid hydrogen one might consider a material with a high 
hydrogen content that was solid at room temperature like polyethylene of 
formula $CH_2$ or lithium hydride of formula $LiH$.

For lithium hydride we have
\begin{equation}\label{hp.1}
0.043/8 = 5 \times 10^{-3}
\end{equation}

and for polyethylene
\begin{equation}\label{hp.2}
2 \times 0.043/14 = 6 \times 10^{-3}.
\end{equation}

If one were to use balls of these materials for a determination of the 
gravitational constant through the Cavendish experiment one should thus get 
lower values for the gravitational constant $G$. In the fifties the value of 
$G$ was uncertain by about 0.1\,\% \cite{AHC} and so the effect might just be
measurable. When one of us (ELS) talked in 1958 to Robert Dicke of Princeton
University about a possible experimental test Dicke had his doubts whether it 
could be done since machining homogeneous spheres in those materials might be 
forbidding. However, a year later it became clear that there should be no 
different value for $G$ with balls of hydrogen. Dieter Brill from Princeton
who had joined the Hamburg seminar alerted us to a paper in the Physical Review
by Charles Misner and  Peter Putnam \cite{MP} about active mass (about Peter 
Putnam who died in 1987 see John Wheeler's Memoir \cite{JW}.) 

Misner and Putnam showed, assuming gravity to be negligible,
 that the 3p-term for a gas in a container was canceled by negative 
contributions to the mass from the stresses in the walls of the container that
kept the gas together. It had not been clear to us that negative 
\emph{surface} contributions to the energy would exactly cancel the positive 
\emph{3p-volume} contribution to the total mass when we had the model of a 
bubble in mind.

\section{The spherical bubble}\label{S:sb}
\setcounter{equation}0

We imagined a gas of constant density $\rho$ and pressure $p$ enclosed in a 
spherical two-dimensional shell of radius $r$ with surface mass density 
$\sigma$ and surface tension $\tau$.
 \bigskip

\begin{figure}
\includegraphics{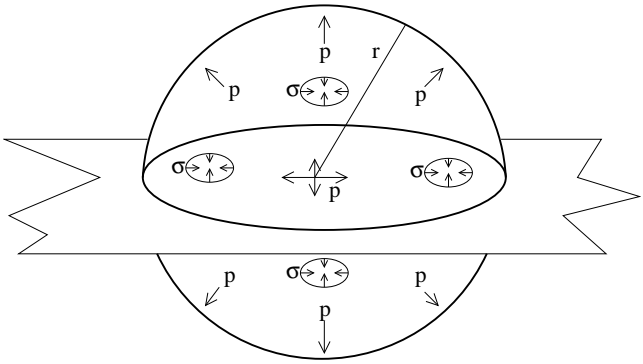}
\caption{A spherical bubble of radius $r$ is filled with a gas of pressure $p$.
The bubble is kept in equilibrium by a surface tension $\sigma$ with dimension
force by length.}\label{Fi:1.}
\end{figure}

\bigskip

The surface tension should be just strong enough for keeping the bubble in
equilibrium. The gravitational binding energy of the bubble was supposed to be
negligible compared to its mass. For finding the relation of surface tension
to pressure for equilibrium we imagined a plane cut through the bubble 
removing the southern hemisphere. To keep the northern hemisphere in 
equilibrium one now had to balance the upward pressure over the equatorial 
disc of
area $\pi r^2$ against the surface tension pulling down along the equator over
the length $2\pi r$. This gives the relation

\bigskip

\begin{figure}
\includegraphics{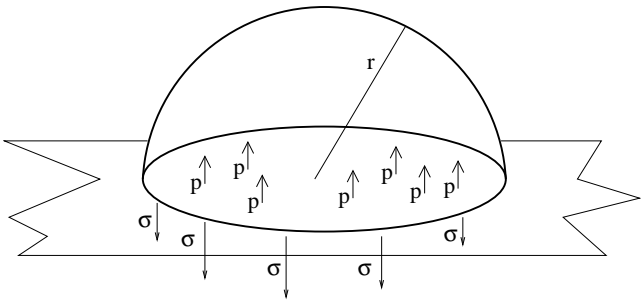}
\caption{The lower hemisphere of the bubble (Fig.\ref{Fi:1.}) is removed and 
replaced by its forces on the upper hemisphere.}\label{Fi:2}
\end{figure}

\bigskip

\begin{equation}\label{sb.1}
p \cdot \pi r^2 = \tau \cdot 2\pi r, \quad \tau = \frac{1}{2}p\,r.
\end{equation}

The total active mass $M$ would then be obtained by the surface contribution 
$4\pi r^2(\sigma - \tau)$ and by the volume contribution 
$4\pi r^3(\rho + 3p)/3$
\begin{equation}\label{sb.2}
M = 4\pi r^2(\sigma - \tau) + \frac{4\pi}{3}r^3(\rho + 3p) =  4\pi r^2\sigma + \frac{4\pi}{3}r^3\rho + 2\pi r^3 p
\end{equation}

leaving us with half the volume contribution of the $3p$-term to the active 
mass.
Something was wrong. It was only last summer in a nostalgic moment when we 
talked about this problem again that we saw the solution:

If the active mass density of a 3-dimensional distribution has to be 
complemented by a $3p$-term, then that of a 2-dimensional shell needs a 
$2p$-term and a 1-dimensional disk a $p$-term (corresponding to the trace of a
2- or 1-dimensional isotropic stress tensor) If they were stresses instead of 
pressures they would come in with a negative sign.

Now all was clear: the surface contribution to the active mass of the bubble
was $4\pi r^2(\sigma - 2\tau)$ and we obtain now instead (\ref{sb.2})
\begin{equation}\label{sb.3}
M = 4\pi r^2(\sigma - 2\tau) + \frac{4\pi}{3}r^3(\rho + 3p) =  4\pi r^2\sigma + \frac{4\pi}{3}r^3\rho .
\end{equation}

This simple remark settled also the case of the active mass of a circular 
disk.

\section{The active mass of a circular disk}\label{S:cd}
\setcounter{equation}0

We consider a circular disk of radius $r$ with mass density $\sigma$ and 
pressure $p$

\bigskip

\begin{figure}
\includegraphics{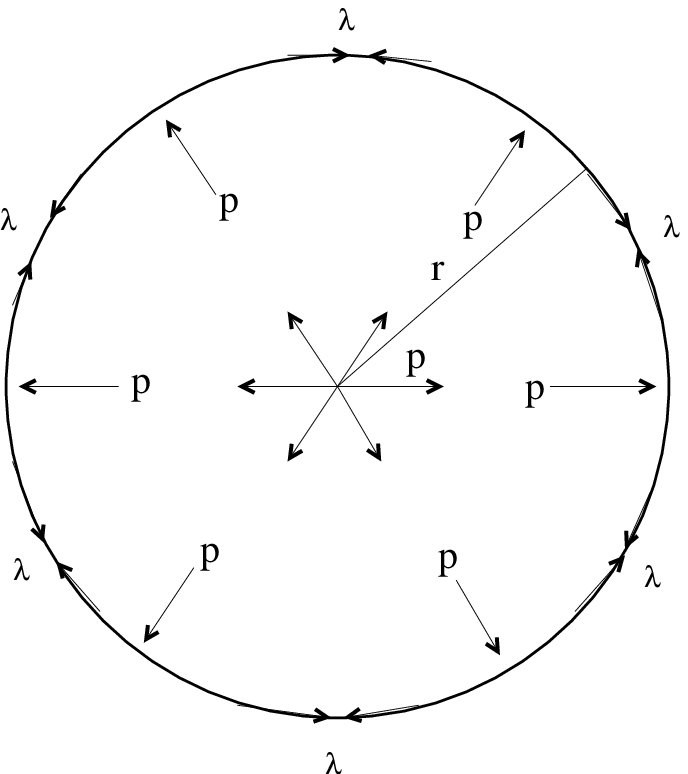}
\caption{Circular disk of radius $r$ carries a surface pressure $p$ with 
dimension force/length. The Pressure is balanced by the tension $\lambda$
with dimension force along its perimeter.}\label{Fi:3.}
\end{figure}

\bigskip

The disk is kept in equilibrium by a one-dimensional string around its 
circumference of linear mass density $\mu$ and stress $\lambda$. To find the 
relation between the pressure $p$ and the stress $\lambda$ we imagine a linear 
cut through the center of the disk removing the lower half

\bigskip

\begin{figure}
\includegraphics{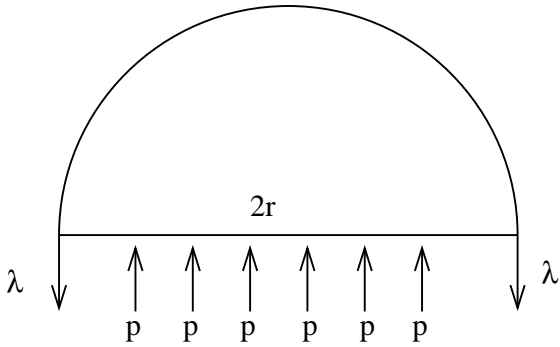}
\caption{The lower half of the disk (Fig. \ref{Fi:3.}) has been removed and replaced by the forces acting on the upper part.}\label{Fi:4.}
\end{figure}

\bigskip

The pressure over the diameter $2r$ must now be balanced against the stress 
$\lambda$ at the left and the right end of the semicircle. This gives
\begin{equation}\label{cd.1}
2r\,p = 2\lambda, \quad \lambda = r\,p.
\end{equation}

The active mass $M$ of the disk is then obtained by taking the active mass 
density $\sigma + 2p$ over the area $\pi r^2$ and adding the active mass 
density $\mu - \lambda$ of the bounding string along the circumference 
$2\pi r$. This gives
\begin{equation}\label{cd.2}
M = (\sigma + 2p)\pi r^2 + (\mu - \lambda)2\pi r = \sigma\pi r^2 + \mu \cdot 2\pi r,
\end{equation}

the promised result.

\section{Conclusion}\label{S:con}
\setcounter{equation}0

In Newtonian theory there was always the understanding, because of ``actio = 
reactio'', that the active 
and passive masses were equal. If this were not the case a system of two 
passive unit masses but different active masses would show an acceleration of
its center of mass violating Newton's law \emph{actio = reactio}. In Einstein's
theory of gravitation the equality of active and passive mass is not so obvious
 since Einstein's cosmological constant can be 
seen as giving rise to self-acceleration for the center of mass of a double 
star. While we have no 
reason to doubt the equality of the three kinds of masses we believe that tests
involving the active mass are certainly desirable. This is especially true for
situations where the gravitational binding energy significantly contributes to
the mass.

In General Relativity the fundamental variables are ``local'' ones,
such as $g_{\alpha\beta}$, $T^{\alpha\beta}$, $\rho$, $u^\alpha$, $p$ \ldots
neither the active nor the passive or the inertial mass of an (extended) body
have been exactly defined so far, these concepts belong to perturbative 
theories in General Relativity.

\section{Killing's Equation}\label{S:a1}
\setcounter{equation}0

A stationary gravitational field is characterized by the existence of a 
time-like Killing field that generates an infinitesimal transformation that 
leaves the metric unchanged. The field is called static if the Killing vector
field is hypersurface-orthogonal; this means the covariant vector is a product 
of a gradient by a scalar function. A Killing vector field $\xi_\mu$ fulfills
the equation
\begin{equation}\label{a1.1}
\xi_{\mu;\nu} + \xi_{\nu;\mu} = 0. 
\end{equation}

If we write down equation (\ref{gr.5}) three times with cyclic permutation of 
the indices
\begin{equation}\label{a1.2}
\xi_{\mu;\alpha;\nu} - \xi_{\mu;\nu;\alpha} = - \xi_\gamma R^\gamma{}_{\mu\alpha\nu},
\end{equation}
\begin{equation}\label{a1.3}
\xi_{\alpha;\nu;\mu} - \xi_{\alpha;\mu;\nu} = - \xi_\gamma R^\gamma{}_{\alpha\nu\mu},
\end{equation}
\begin{equation}\label{a1.4}
\xi_{\nu;\mu;\alpha} - \xi_{\nu\alpha;\mu} = - \xi_\gamma R^\gamma{}_{\nu\mu\alpha},
\end{equation}

adding the first two equations and subtracting the third, we have because of
(\ref{a1.1}) and the cyclic symmetry of the Riemann tensor
\begin{equation}\label{a1.5}
R^\gamma{}_{\mu\alpha\nu} + R^\gamma{}_{\alpha\nu\mu} + R^\gamma{}_{\nu\mu\alpha} = 0
\end{equation}

that
\begin{equation}\label{a1.6}
2\,\xi_{\mu;\alpha;\nu} = 2\,\xi_\gamma R^\gamma{}_{\nu\mu\alpha}.
\end{equation}

These equations are known as the integrability conditions for the Killing 
field. If we define
\begin{equation}\label{a1.7}
\xi_{\mu;\alpha} - \xi_{\alpha;\mu} = 2\,\xi_{\mu;\alpha} \equiv F_{\mu\alpha} \quad \text{and} \quad 2\,\xi_\gamma R^\gamma{}_\mu \equiv j_\mu,
\end{equation}

we obtain from (\ref{a1.6}) by anti-symmetrization and contraction Maxwell's 
equations for a field tensor $F_{\mu\nu}$ and a four-current $j_\mu$
\begin{equation}\label{a1.8}
F_{\mu\alpha;\nu} + F_{\alpha\nu;\mu} + F_{\nu\mu;\alpha} = 0, \quad F^{\mu\nu}{}_{;\nu} = j^\mu.
\end{equation}

Since
\begin{equation}\label{a1.9}
\xi_{\mu;\alpha} - \xi_{\alpha;\mu} = \xi_{\mu,\alpha} - \xi_{\alpha,\mu} = F_{\mu\alpha}
\end{equation}

the Killing vector for a stationary gravitational field plays the r\^{o}le of
an electromagnetic four-potential.

\section{Adapted Coordinates}\label{S:a2}
\setcounter{equation}0

We choose coordinates in such a way that
\begin{equation}\label{a2.1}
\xi^\mu = \delta^\mu_0.
\end{equation}

That can be done for any contravariant vector field in a finite region. This 
specialization still allows gauge transformation with an arbitrary function
$\chi(x^k)$
\begin{equation}\label{a2.2}
\bar{x}^0 = x^0 + \chi(x^k), \quad d\bar{x}^0 = dx^0 + \chi_{,j}\,dx^j, \quad \bar{x}^j = \bar{x}^j(x^k).
\end{equation}

The Killing equation (\ref{a1.1}) can be written
\begin{equation}\label{a2.3}
g_{\mu\nu,\lambda}\xi^\lambda + g_{\lambda\nu}\xi^\lambda{}_{,\mu} + g_{\mu\lambda}\xi^\lambda{}_{,\nu} = 0.
\end{equation}

This gives with our normalization (\ref{a2.1})
\begin{equation}\label{a2.4}
g_{\mu\nu,0} = 0,
\end{equation}

that is independence of the coordinate $x^0$. Since we requested that the 
Killing vector $\xi^\mu$ be time-like we have that $x^0$ is a distinguished 
time coordinate with
\begin{equation}\label{a2.5}
\xi_\mu = g_{0\mu}, \quad \xi^\mu\xi_\mu = g_{00} > 0.
\end{equation}

Using now the condition that the Killing vector is hypersurface-orthogonal 
(static field) 
\begin{equation}\label{a2.6}
\xi_\mu(\xi_{\nu,\lambda} - \xi_{\lambda,\nu}) + \xi_\nu(\xi_{\lambda,\mu} - \xi_{\mu,\lambda}) + \xi_\lambda(\xi_{\mu,\nu} - \xi_{\nu,\mu}) = 0
\end{equation}

we have after contraction with $\xi^\mu$,
\begin{equation}\label{a2.7}
g_{00}(g_{0\nu,\lambda} - g_{0\lambda,\nu}) - g_{0\nu}\,g_{00,\lambda} + g_{0\lambda}\,g_{00,\nu} = 0.
\end{equation}

This gives after division by $(g_{00})^2$
\begin{equation}\label{a2.8}
(g_{0\nu}/g_{00})_{,\lambda} - (g_{0\lambda}/g_{00})_{,\nu} = 0.
\end{equation}

We have, therefore, that $g_{0\nu}/g_{00}$ is a gradient of a scalar function 
$\psi$
\begin{equation}\label{a2.9}
g_{0\nu} = g_{00}\psi_{,\nu},   \quad   \psi_{,0} = 1.
\end{equation}

We can write
\begin{equation}\label{a2.10}
d\psi = dx^0 + \psi_{,j}(x^k)\,dx^j.
\end{equation}

Comparison with (\ref{a2.2}) shows that we can choose the gauge transformation
$\chi$ such that
\begin{equation}\label{a2.11}
\psi = \bar{x}^0.
\end{equation}

By dropping the bar on $\bar{x}^0$ we have then for the metric the form 
(\ref{gr.8}) 
\begin{equation}\label{a2.12}
ds^2 = g_{00}(x^l)\,(dx^0)^2 + g_{jk}(x^l)\,dx^jdx^k.
\end{equation}

The purely spatial coordinate transformations are still free. The time-like 
hypersurface-orthogonal Killing vector is, in general, unique up to a constant
factor. The square of this factor multiplies $g_{00}$.

\section{Poisson Equation}\label{S:a3}
\setcounter{equation}0

The length of the Killing vector is $\sqrt{g_{00}}$. A state of rest is then
desribed by a time-like unit vector
\begin{equation}\label{a3.1}
u^\mu = \frac{1}{\sqrt{g_{00}}}\xi^\mu, \quad u^\mu u_\mu = 1.
\end{equation}

We study now the second set of Maxwell's equations (\ref{a1.8}) in adapted 
coordinates. We have because of time independence
\begin{equation}\label{a3.2}
F^{\mu\nu}{}_{;\nu} = \frac{1}{\sqrt{-g}}\Big(\sqrt{-g}F^{\mu\nu}\Big)_{,\nu} = \frac{1}{\sqrt{-g}}\Big(\sqrt{-g}F^{\mu k}\Big)_{,k}. 
\end{equation}

Since by (\ref{a1.7})
\begin{equation}\label{a3.3}
F_{\mu\nu} = \Big(g_{00} \delta^0_\mu\Big)_{,\nu} - \Big(g_{00} \delta^0_\nu\Big)_{,\mu},
\end{equation}

the only non-vanishing covariant components are
\begin{equation}\label{a3.4}
F_{0j}  = g_{00,j} = - F_{j0}.
\end{equation}

Entered into (\ref{a3.2}) this gives
\begin{equation}\label{a3.5}
F^{\mu\nu}{}_{;\nu} = \delta^\mu_0\frac{1}{\sqrt{-g}}\Big(\sqrt{-g}\,g_{00,j}\,\frac{1}{g_{00}}\,g^{jk}\Big)_{,k}
\end{equation} 

with
\begin{equation}\label{a3.6}
\sqrt{-g} = \sqrt{g_{00}}\sqrt{-\hat{g}}, \quad \hat{g} = \det \parallel g_{jk}\parallel.
\end{equation} 

We obtain
\begin{equation}\label{a3.7}
F^{\mu\nu}{}_{;\nu} = \delta^\mu_0\frac{2}{\sqrt{-g}}\Big(\sqrt{-\hat{g}}\,g^{jk}(\sqrt{g_{00}})_{,j}\Big)_{,k} = - \frac{2}{\sqrt{g_{00}}}\delta^\mu_0\bigtriangledown^2\sqrt{g_{00}}.
\end{equation} 

The four-current $j_\mu$ from (\ref{a1.7}) becomes with the Einstein field 
equations (\ref{gr.4})
\begin{equation}\label{a3.8}
j_\mu = 2\,\xi_\gamma R^\gamma{}_\mu = 2\Big(-\kappa T^\gamma{}_\mu + \frac{1}{2}\,\delta^\gamma{}_\mu\,\kappa\,T + \lambda\,\delta^\gamma{}_\mu\Big)\xi_\gamma
\end{equation} 

or
\begin{equation}\label{a3.9}
j^\mu = 2\Big(- \kappa\,T^\mu_0 + \frac{1}{2}\,\delta^\mu_0\,(\kappa\,T + 2\lambda)\Big).
\end{equation} 

Here $T$ is the trace (contraction) of the tensor $T^\mu{}_\nu$. The second set 
of Maxwell's equations state that
\begin{equation}\label{a3.10}
T_0{}^j = 0
\end{equation} 

saying that the density of momentum and energy flux vanish in this static 
situation. Calling the energy density $\rho$ and the trace of the pressure
tensor $3p$
\begin{equation}\label{a3.11}
T = \rho - 3p, \quad \rho \equiv T^0{}_0, \quad 3p = - T^a{}_a
\end{equation} 

we have
\begin{equation}\label{a3.12}
j^0 = -\kappa\,(\rho + 3p) + 2\lambda.
\end{equation} 

With (\ref{a3.7}) this gives the relativistic Poisson equation for a static 
gravitational field.
\begin{equation}\label{a3.13}
\bigtriangledown^2\sqrt{g_{00}} = \sqrt{g_{00}}\Big(4\pi(\rho + 3p) - \lambda\Big).
\end{equation}

\section{The energy of a particle in a static gravitational field}\label{S:eps}
\setcounter{equation}0

A particle of constant mass $m$ and four-velocity $v^\mu$ moving on a geodesic
in a static gravitational field obeys
\begin{equation}\label{eps.1}
v^\mu v_\mu = 1, \quad \dot{v}^\mu \equiv v^\mu{}_{;\nu} v^\nu = 0.
\end{equation}

The energy integral for unit mass is given by
\begin{equation}\label{eps.2}
E = v^\mu\,\xi_\mu
\end{equation}

since
\begin{equation}\label{eps.3}
\dot{E} = (v^\mu\,\xi_\mu)\spdot =  v^\mu{}_{;\nu} v^\nu\,\xi_\mu + \xi_{\mu;\nu} v^\mu\,v^\nu = 0.
\end{equation}

The first term on the right hand side vanishes because of the geodesic equation
(\ref{eps.1}) while the second term is zero due to the Killing equation 
(\ref{a1.1}).

In terms of the components of the local rest frame we have
\begin{equation}\label{eps.4}
E = \frac{m}{\sqrt{1-\beta^2}}\, \sqrt{g_{00}}
\end{equation}

where $\beta$ is the local velocity in terms of the speed of light.

\section{Acceleration of a particle at rest}\label{S:apr}
\setcounter{equation}0

A particle at rest is characterised by its four-velocity
\begin{equation}\label{apr.1}
u_\mu = \frac{1}{\sqrt{g_{00}}}\,\delta^0_\mu
\end{equation}

The acceleration of the particle is given by
\begin{equation}\label{apr.2}
\dot{u}_\mu \equiv u_{\mu;\nu}\,u^\nu = u_{\mu,\nu}\,u^\nu - \Gamma_{\mu\nu}{}^\lambda u_\lambda u^\nu
\end{equation}

Since $g_{00}$ is independent of time the first term on the right-hand side
vanishes. We obtain then with (\ref{apr.1})
\begin{equation}\label{apr.3}
\dot{u}_\mu = - \Gamma_{\mu\nu}{}^\lambda u_\lambda u^\nu = - \frac{1}{g_{00}}\,\Gamma_{\mu 0,0} = - \frac{1}{2\,g_{00}}\,g_{00,\mu}
\end{equation}

or
\begin{equation}\label{apr.4}
\dot{u}_k = - \frac{1}{2\,g_{00}}\,g_{00,k} = - \frac{(\sqrt{g_{00}})_{,k}}{\sqrt{g_{00}}}.
\end{equation}

\section{The pressure of an ideal Fermi gas at zero temperature}\label{S:b1}
\setcounter{equation}0

The pressure for such a gas at zero temperature was derived by Enrico Fermi in
his 1926 paper ``On the quantization of the ideal mono-atomic gas'' \cite{EF}.
For $N$ identical particles in volume $V$ we have a number density $n = N/V$
and according to (\ref{tp.1})
\begin{equation}\label{b1.1}
p\,V = \frac{1}{3}\,N\,\overline{\vec{p} \cdot \vec{v}}.
\end{equation}

For a non-relativistic motion this gives with a mass $m$
\begin{equation}\label{b1.2}
p\,V = \frac{N}{3}\,\overline{(\vec{p})^2}/m = \frac{2}{3}N\,\bar{\epsilon}
\end{equation}

where $\bar{\epsilon}$ is their average kinetic energy.

For an ideal Fermi gas at zero temperature in a cubic box of side length $L$ 
for a statistical weight $g$ we can fit $g$ particles in a cell of phase 
space of size
\begin{equation}\label{b1.3}
L^3 \cdot \triangle p_x \cdot \triangle p_y \cdot \triangle p_z = h^3.
\end{equation}

While Fermi took $g=1$, the case of general $g$ was discussed by Wolfgang Pauli in his 1927 paper ``On gas degeneration and paramagnetism'' \cite{WP}.

We have thus for $dN$ particles
\begin{equation}\label{b1.4}
dN = g \cdot 4\pi\,(\vec{p})^2 d|\,\vec{p}\,|\Big(\frac{L}{h}\Big)^3
\end{equation}

and for the total number $N$ of particles in terms of the maximum Fermi 
momentum $P_F$
\begin{equation}\label{b1.5}
N = g \frac{4\pi}{3}\Big(\frac{P_F \cdot L}{h}\Big)^3, \quad n = \frac{N}{V} =  g \frac{4\pi}{3}\Big(\frac{P_F}{h}\Big)^3.
\end{equation}

We obtain the average kinetic energy by averaging the expression $(\vec{p})^2/2m$ and obtain
\begin{equation}\label{b1.6}
\bar{\epsilon} = \frac{1}{N}\int \frac{(\vec{p})^2}{2m}dN = \frac{3}{2 m p_F^3}\int\limits^{P_F}_0 (\vec{p})^4 d|\,\vec{p}\,| = \frac{3 P_F^2}{10 m}.
\end{equation}

This gives for the pressure from (\ref{b1.2})
\begin{equation}\label{b1.7}
p = \frac{n}{5m}P_F^2.
\end{equation}

The ratio of $3p/c^2$ to the mass density $\rho$ becomes
\begin{equation}\label{b1.8}
\frac{3p}{\rho c^2} = \frac{3}{5}\Big(\frac{P_F}{m c}\Big)^2
\end{equation}

The equation (\ref{b1.5}) allows us to express the Fermi momentum in terms of
the particle density and we obtain thus
\begin{equation}\label{b1.9}
P_F = h\Big(\frac{3 n}{4\pi g}\Big)^{1/3}.
\end{equation}

Substituting this expression into (\ref{b1.8}) gives
\begin{equation}\label{b1.10}
\frac{3 p}{\rho c^2} = \frac{3}{5}\Big(\frac{h}{m c}\Big)^2\Big(\frac{3n}{4\pi g}\Big)^{2/3}
\end{equation}

or in terms of Dirac's $\hbar$
\begin{equation}\label{b1.11}
\frac{3 p}{\rho c^2} = \frac{3}{5}\Big(\frac{\hbar}{m c}\Big)^2\Big(\frac{6\pi^2 n}{g}\Big)^{2/3}.
\end{equation}

\section{The atomic nucleus as a degenerate Fermi gas}\label{S:b2}
\setcounter{equation}0
 
Werner Karl Heisenberg considered an atomic nucleus as a Fermi gas of nucleons
in his 1933 paper \cite{WKH} ``On the Structure of atomic nuclei III''. We want
to calculate the term $3p/c^2\rho$ for an ideal Fermi gas of nucleons at zero
temperature.The interactions among nucleons that keep them confined are simply
taken into account by a potential well in which they move freely. For a rough 
estimate we disregard the difference between protons and neutrons and take the 
statistical weight $g$ to be
\begin{equation}\label{b2.1}
g = 4.
\end{equation}

For a nucleus with $N$ nucleons we take the density of nuclear matter
\begin{equation}\label{b2.2}
n = \frac{N}{N \cdot (4\pi/3)r_0^3} = \frac{3}{4\pi r_0^3}.
\end{equation}

The constant $r_0$ is given by \cite{KH}
\begin{equation}\label{b2.3}
r_0 = 1.2\,fm.
\end{equation}

We have then from (\ref{b1.11})
\begin{equation}\label{b2.4}
\frac{3p}{\rho c^2} = \frac{3}{5}\Big(\frac{\hbar}{m c r_0}\Big)^2\Big(\frac{9\pi}{8}\Big)^{2/3}.
\end{equation}

This gives
\begin{equation}\label{b2.5}
\frac{3p}{\rho c^2} = 1.39\Big(\frac{\hbar}{m c r_0}\Big)^2 = 0.043.
\end{equation}

We thank Yanwen Shang and Eugene Surowitz for comments and Linda Snow for
locating some of the more arcane references.

\newpage

Fig 1. A spherical bubble of radius $r$ is filled with a gas of pressure $p$. 
The bubble is kept in equilibrium by a surface tension $\sigma$ with dimension 
force by length.

Fig 2. The lower hemisphere of the bubble of Fig. 1. is removed and replaced 
by its forces on the upper hemisphere.

Fig 3. A circular disk of radius $r$ carries a surface pressure $p$ with 
dimension force/length. The pressure is balanced by the tension $\lambda$ with
dimension force along its perimeter.

Fig 4. The lower half of the disk in Fig. 3. has been removed and replaced by 
the forces acting on the upper part.

\newpage

\begin{center}
\textbf{The authors}
\end{center}

\begin{center}
J\"{u}rgen Ehlers \\
Max-Planck-Institut f\"{u}r Gravitationsphysik, Golm, Germany \\
Istv\'{a}n Ozsv\'{a}th \# \\
Department of Mathematics, The University of Texas at Dallas, \\
Richardson, Texas 75083-0688 \\
Engelbert L. Schucking  \\
Department of Physics, New York University, 4 Washington Place, \\
New York 10003  \\

\#  Electronic mail: ozsvath@utdallas.edu 
\end{center}

\end{document}